\documentclass[twocolumn,showpacs,preprintnumbers,amsmath,amssymb,superscriptaddress]{revtex4}

\usepackage{graphicx}
\usepackage{dcolumn}
\usepackage{bm}
\usepackage{enumerate}

\newcommand{\diff}{\mathrm{d}}
\begin{document}


\title{Evolving networks by merging cliques}

\author{Kazuhiro Takemoto}
\email{d673050k@bio.kyutech.ac.jp}
\affiliation{
Department of Bioscience and Bioinformatics, Kyushu Institute of Technology, Iizuka Fukuoka 820-8502, Japan
}
\author{Chikoo Oosawa}%
\email{chikoo@bio.kyutech.ac.jp}
\affiliation{
Department of Bioscience and Bioinformatics, Kyushu Institute of Technology, Iizuka Fukuoka 820-8502, Japan
}
\affiliation{
Bioalgrithm Project, Faculty of Computer Science and Systems Engineering, Kyushu Institute of Technology, Iizuka Fukuoka 820-8502, Japan
}

\date{October 7, 2005}

\begin{abstract}
We propose a model for evolving networks by merging building blocks represented as complete graphs, reminiscent of modules in biological system or communities in sociology. The model shows power-law degree distributions, power-law clustering spectra and high average clustering coefficients independent of network size. The analytical solutions indicate that a degree exponent is determined by the ratio of the number of merging nodes to that of all nodes in the blocks, demonstrating that the exponent is tunable, and are also applicable when the blocks are classical networks such as Erd\H{o}s-R\'enyi or regular graphs. Our model becomes the same model as the Barab\'asi-Albert model under a specific condition.
\end{abstract}

\maketitle

\section{Introduction}
Complex networks, evolved from the Erd\H{o}s-R\'enyi (ER) random network \cite{ref:ERmodel}, are powerful models that can simply describe  complex systems in many fields such as biology, sociology, and ecology, or information infrastructure,
World-Wide Web, and Internet \cite{book:evo-net,book:internet,ref:network-review,ref:network-bio}. In particular, some striking statistical properties in real-world complex networks have been revealed in recent years. The network models reproduce the properties and promise to understand growth and control mechanisms of the systems.

One of the striking properties in the real-world networks is a scale-free feature: power-law degree distributions are defined as existence probability of nodes with degree (number of edges) $k$; $P(k)\sim k^{-\gamma}$ with $2<\gamma<3$ are empirically found \cite{book:evo-net,book:internet,ref:network-review,ref:SF}. The feature can not be explained by the ER model because the model shows Poisson distribution. However, the Barabasi-Albert (BA) model \cite{ref:SF,ref:BAmodel} exhibits power-law degree distributions. The model is well known as a scale-free network model and consists of the two mechanisms: growth and preferential attachment,
\begin{equation}
\Pi_i=\frac{k_i}{\sum_jk_j}
\label{eq:PA}
\end{equation}
denotes the probability that node $i$ is chosen to get an edge from the new node, and is proportional to degree of node $i$; $k_i$. Equation (\ref{eq:PA}) means that high-degree nodes get an even better chance to attract next new edges; the rich get richer. The model indicates that $P(k)\sim k^{-3}$ and the degree exponent is fixed. After that, extended BA models with modified preferential attachments, including weight \cite{ref:weight} or competitive \cite{ref:fitness} dynamics, and/or local rules \cite{ref:local-event}, rewiring, and adding of edges, are proposed to reproduce statistical properties between BA model networks and real-world networks. In addition, exponential-like distributions are often observed in real-world networks \cite{ref:Amaral-2000,ref:Sen-2003}. The distributions are reproduced by an extended BA model with aging and saturation effects \cite{ref:Amaral-2000}, nonlinear preferential attachment rule \cite{ref:Barabasi-2002}, or controllability of growth and preferential attachment \cite{ref:Shargel-2003}.

The other of the striking properties is a small-world feature: significantly high clustering coefficients $C$ denote density of edges between neighbors of a given node and imply clique (cluster) structures in
the networks \cite{ref:small-world}. The structures correspond to communities in social networks and network motifs \cite{ref:network-motif} such as feedforward and feedback loops in biological and technological networks. Emergence of the clique structures in the networks are called "transitivity phenomena" \cite{ref:trans}.

In recent years, the transitivity in the many networks are actively investigated with statistical approaches and it is found that power-law clustering spectra are defined as correlations between degree $k$ of a given node and the clustering coefficient $C$ of the node; $C(k)\sim k^{-\alpha}$ with $\alpha\leq 1$ in rough is found in the numerical analyses \cite{ref:metabo-module,ref:determin,ref:local-strategy}.  More specifically, $\alpha$ exhibits around 1, suggesting a hierarchical structure of the cliques \cite{ref:metabo-module,ref:determin}.

In modeling approaches for the structures, the extended BA models with aging \cite{ref:aging} or triad formation \cite{ref:triad_01,ref:triad_02} and Ravasz's hierarchical model \cite{ref:determin} have been proposed because of the absence of the structure in original BA networks.  In particular, the hierarchical model evolves determinably with replication of complete graphs as cliques, providing a power-law clustering spectrum; $C(k)\sim k^{-1}$ and degree distribution with arbitrary degree exponent. The model takes into account systematic reorganization of cliques as functional modules or communities and the consideration is important for understanding developmental processes and controls in the systems.

In this paper, we propose that an evolving network model with reorganization of cliques is constitutional unit (basic building block). The model is inspired by the elimination of deterministic growing process in Ravasz's hierarchical model, providing high general versatility for growing mechanisms. Moreover, the model characterizes a relationship between statistical properties and compositions of the cliques.

We explain details of the model in Sec. \ref{sec:model}, and the analytical solutions with mean-field continuous approaches of the statistical properties in Sec. \ref{sec:ana}, the comparisons between the numerical and the analytical solutions in Sec. \ref{sec:num}, and conclude this paper in Sec. \ref{sec:conc}.

\section{Model}
\label{sec:model}
\begin{figure}[ht]
\begin{center}
	\includegraphics{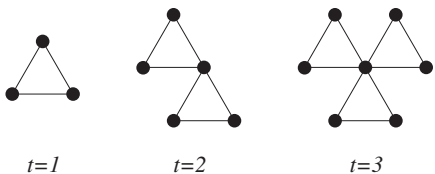}  
	\caption{Schematic diagram of growth process of the model network with $a=3$ and $m=1$. Each clique is merged through common node(s) without adding extra edges.}
	\label{fig:MM}
\end{center}
\end{figure}

Here we present an evolving network model that the mechanism has the following three procedures (see Fig. \ref{fig:MM}): 
\begin{enumerate}[i)]
\item We start from a clique as a complete graph with $a(>2)$ nodes.
\item At every time step the new clique with the same size is joined by merging to existing $m(<a)$ node(s). Please note that cliques are merged {\em without} adding extra links.
\item When merging the graphs, the preferential attachment (PA) rule, Eq. (\ref{eq:PA}), is used to select $m$ old nodes and resultant duplicated edge(s) between the merged nodes are counted and contribute to PA in the next time step.  [Please imagine that all edges in the clique are stretchable, then a node in the clique can reach any existing nodes. Any old nodes can be targeted by node(s) in the new clique.]
\end{enumerate}
With the $a>m$ condition, networks grow in time steps.

\section{Analytical solutions}
\label{sec:ana}
\subsection{Degree distribution}
\label{subsec:deg-ana}
 The degree distribution is defined as the existence probability of nodes with degree $k$, and is formulated as
\begin{equation}
P(k)=\frac{1}{N}\sum_{i=1}^N\delta(k_i-k),
\label{eq:def-deg}
\end{equation}
where $\delta(x)$ is Kronecker's delta function. To describe the degree distribution of our model we take the continuous mean-field approach used by many authors \cite{book:evo-net,book:internet,ref:BAmodel}. Since a network in our model evolves in every clique, the standard approach can not be applied directly.

We take the following method called the coarse-graining approach to be applied to the standard continuous mean-field approach.
\begin{figure}[ht]
\begin{center}
	\includegraphics{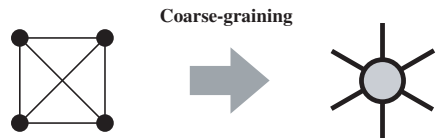} 
\end{center}
	\caption{Schematic diagram of coarse-graining method ($a=4$).}
	\label{fig:henkan}
\end{figure}
Let the $a$-size clique be regarded as a grain with ${a \choose 2}$ edges (see Fig. \ref{fig:henkan}). Then, edges connecting to $m$ merged nodes in the clique can be considered as edges join to other grains, and ${a-m \choose 2}$ edges in the clique are futile or do not link to the other grains. That is, the relationship between $G_i$ are the degree of a grain and $k_i$ are expressed as
\begin{equation}
G_i=k_i+\kappa_0,
\label{eq:G-k}
\end{equation}
where $\kappa_0$ corresponds to ${a-m \choose 2}$.

Now the standard approach can be applied; a time evolution of the degree of a $G_i$ can be written as
\begin{equation}
\frac{{\diff}G_i}{{\diff}t}=m(a-1)\frac{G_i}{\sum_jG_j},
\label{eq:diff-G}
\end{equation}
where $\sum_jG_j=2{a \choose 2}t$. The solution of the equation with $G_i(t=s)={a \choose 2}$ as an initial condition for Eq. (\ref{eq:diff-G}) is 
\begin{equation}
G_i(t)={a \choose 2}\left(\frac{t}{s}\right)^{\rho},
\label{eq:Gevo}
\end{equation}
where
\begin{equation}
\rho=m/a
\label{eq:ratio}
\end{equation}
represents the ratio between the number of merged node(s) and that of all nodes in the clique.  Please note that Eq. (\ref{eq:ratio}) also satisfies the case that the clique is a regular graph or a random graph \cite{ref:ERmodel} because the graphs have homogeneous degrees as well as complete graphs.

By using the continuous approach, the probability distribution of $G_i$ can be obtained,
\begin{equation}
P(G)=\frac{{a \choose 2}^{1/\rho}}{{\rho}G^\gamma}.
\label{eq:P(G)}
\end{equation}
Substituting Eq. (\ref{eq:G-k}) into Eq. (\ref{eq:P(G)}) we obtain
\begin{equation}
P(k)={\cal A}(a,\rho)(k+\kappa_0)^{-\gamma},
\label{eq:deg}
\end{equation}
where ${\cal A}(a,\rho)={a \choose 2}^{1/\rho}/{\rho}$, demonstrating that the distribution has power-law fashion and cutoff in lower degrees, and the exponent is 
\begin{equation}
\gamma=\frac{\rho+1}{\rho}.
\label{eq:gamma}
\end{equation}
Equation (\ref{eq:gamma}) shows a direct relationship between the exponent of the distribution and the ratio shown Eq. (\ref{eq:ratio}).  To establish the correspondence of our model to the BA model \cite{ref:BAmodel}, one can assign 2 to $a$ and 1 to $m$; we can calculate that the exponent in the condition is $(0.5+1)/0.5=3$, showing that our model can be recognized as a more expanded model than BA.

\subsection{Clustering spectrum}
\label{subsec:clus-ana}
It is well known that complex networks have other statistical properties.  Next we step into analytical treatment of a clustering spectrum that can estimate characteristics of the hierarchy of networks' modularity. the clustering spectrum is defined as 
\begin{equation}
C(k)=\frac{1}{NP(k)}\sum_{i=1}^NC_i\times\delta(k-k_i),
\label{eq:clus-spec}
\end{equation}
where $\delta(x)$ is Kronecker's delta function, and $C_i$ denotes the clustering coefficient defined by 
\begin{equation}
C_i=\frac{M_i}{{k_i\choose 2}}=\frac{2M_i}{k_i(k_i-1)},
\label{eq:clus}
\end{equation}
and means density of neighboring edges of the node, where $k_i$ and $M_i$ denote the degree of node $i$ and the number of edges between the neighbors, respectively.

\begin{figure}[ht]
\begin{center}
	\includegraphics{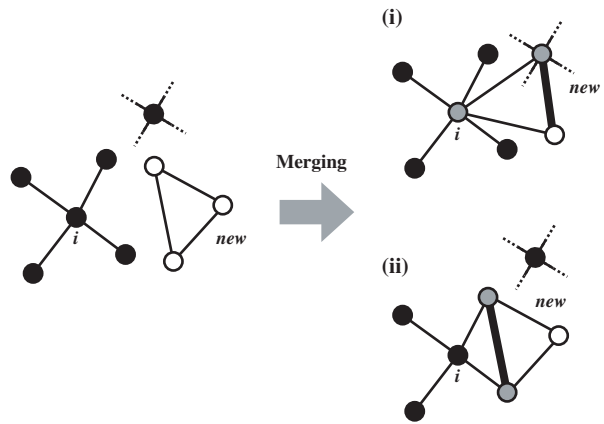} 
\end{center}
	\caption{Conditions for increasing of $M_i$ ($a=3$, $m=2$). The existing nodes are filled with black, the new nodes are open circles, and the merged nodes are filled with gray. The thick lines are edges between nearest neighbors of node $i$.}
	\label{fig:clus_pro}
\end{figure}

To derive the analytical solution of Eq. (\ref{eq:clus-spec}), we need to obtain the formula of $M_i$ first. The following two conditions that $M_i$ increases are need to be examined:
\begin{enumerate}[(i)]
\item The condition that node $i$ is merged to a node of a clique as a complete graph, and the other node(s) of the clique are merged to existing nodes [see Fig. \ref{fig:clus_pro} (i)].
\item The condition that the new clique is merged to node(s) that are neighboring to node $i$ [see Fig. \ref{fig:clus_pro} (ii)].
\end{enumerate}

Since both conditions independently contribute to an increase of $M_i^{\mathrm{(i)}}$, $M_i^{\mathrm{(ii)}}$ can be expressed as the sum of both effects,
\begin{equation}
M_i=M_{i}^{\mathrm{(i)}}+M_{i}^{\mathrm{(ii)}}.
\end{equation}
Because we assume that a clique is a complete graph, condition (i) becomes $M_i^{\mathrm{(i)}}$ concrete,
\begin{equation}
M_i^{\mathrm{(i)}}=\frac{a-2}{2}k_i.
\end{equation}

\begin{figure}[ht]
\begin{center}
	\includegraphics{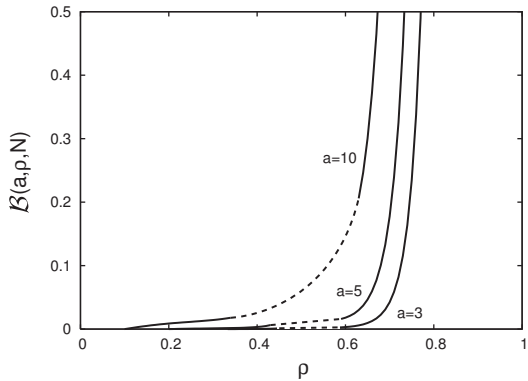} 
\end{center}
	\caption{The dependence of ${\cal B}(a,\rho,N)$ on $\rho$ with $N=50,000$.  With larger $\rho$ and/or $a$ uncertainty to the distribution of clustering coefficient $C(k)$ becomes larger.}
	\label{fig:B}
\end{figure}

$M_i^{\mathrm{(ii)}}$ increases when neighboring nodes to node $i$ are consecutively chosen by the PA rule. In other words, $M_i^{\mathrm{(ii)}}$ is proportional to degrees of the neighboring nodes.
By using the average nearest-neighbor degree of node $i$, $\langle k_{nn}\rangle_i$, we can write the rate equation of $M_i^{\mathrm{(ii)}}$ with the continuous approach \cite{ref:triad_02},
\begin{equation}
\frac{{\diff}M_{i}^{\mathrm{(ii)}}}{{\diff}t}={m \choose 2}{k_i \choose 2}\left(\frac{\langle k_{nn} \rangle_i}{\sum_jk_j}\right)^2.
\label{eq:M}
\end{equation}
To go to further analytical treatment, both analytical and numerical results help to simplify Eq. (\ref{eq:M}). $\langle k_{nn}\rangle_i$ can be expressed by using degree correlation \cite{ref:degree-correlation} $\bar{k}_{nn}(k)$ denoting that correlations between nodes with $k$ degree and the nearest-neighbors degree to the nodes. Based on detailed analysis (see the Appendix), we can assume that $\bar{k}_{nn}(k)$ is uncorrelated with $k$, leading to further simplification reported by Egu\'iluz {\em et al.} \cite{ref:knn}. They show that $\langle k_{nn}\rangle=\langle k^2 \rangle/\langle k \rangle$ for uncorrelated networks, where $\langle k^2 \rangle$ and $\langle k \rangle$ means the average of the square of $k$ and that of $k$, respectively. The average degree $\langle k \rangle$ in our model is
\begin{equation}
\langle k \rangle=\frac{a(a-1)}{a-m}.
\label{eq:ave_k}
\end{equation}
With $k_i(t)\simeq{a \choose 2}(t/s)^{\rho}$ given with Eqs. (\ref{eq:Gevo}) and (\ref{eq:G-k}), the average of square of $k$, $\langle k^2 \rangle$, is expressed as
\begin{equation}
\langle k^2 \rangle {\simeq}\frac{1}{(a-m)t}\int_{1}^t\left[{a \choose 2}\left(\frac{t}{s}\right)^\rho\right]^2{\diff}s.
\label{eq:k2}
\end{equation}
Equation (\ref{eq:k2}) represents that $\langle k^2 \rangle$ depends on $t$. With Eq. (\ref{eq:ave_k}), we get the approximation of time evolution of $\langle k_{nn} \rangle$ is sensitive to $\rho$,
\begin{equation} 
\langle k_{nn} \rangle \simeq
\left\{
\begin{array}{ll}
\frac{a(a-1)}{4(1-2\rho)}=\mathrm{const.} & (0<\rho<0.5) \\
\frac{a(a-1)}{4}\ln t & (\rho=0.5) \\
\frac{a(a-1)}{4(2\rho-1)}t^{2\rho-1} & (0.5<\rho<1). \\
\end{array}\label{eq:kk}
\right.
\end{equation}
Substituting this into Eq. (\ref{eq:M}) with the initial condition $M_i^{\mathrm{(ii)}}(t=1)=0$, we obtain following Eq. (\ref{eq:hantei2}) that shows dependence of $M_i^{\mathrm{(ii)}}$ on time,
\begin{equation} 
M_i^{\mathrm{(ii)}}\propto
\left\{
\begin{array}{ll}
k_i^2t^{-2\rho} & (0<\rho<0.5) \\
k_i^2\ln^3t/t & (\rho=0.5) \\
k_i^2t^{4\rho-3} & (0.5<\rho<1). \\
\end{array}
\right.\label{eq:hantei2}
\end{equation}
Finally the analytical solution of clustering spectrum becomes
\begin{equation} 
C(k){\simeq}\frac{a-2}{k}+{\cal B}(a,\rho,N),
\label{eq:cluster}
\end{equation}
where ${\cal B}(a,\rho,N)$ gives positive value (see Fig. \ref{fig:B}) and is expressed as
\begin{equation} 
{\cal B}(a,\rho,N)=
\left\{
\begin{array}{ll}
\frac{a\rho(a\rho-1)}{32(1-2\rho)^3}N_G^{-2\rho} & (0<\rho<0.5) \\
\frac{a(a-2)}{64}\ln^3N_G/N_G & (\rho=0.5) \\
\frac{a\rho(a\rho-1)}{96(2\rho-1)^3}N_G^{4\rho-3} & (0.5<\rho<1), \\
\end{array}
\right.\label{eq:B}
\end{equation}
where $N_G=N/(a-m)$.

Figure \ref{fig:B} is depicted from Eq. (\ref{eq:B}). For smaller $\rho\leq 0.5$ and $a$, ${\cal B}(a,\rho,N)$ take smaller values, yielding distribution of clustering coefficient $C(k)\sim k^{-1}$ from Eq. (\ref{eq:cluster}).  For larger $\rho>0.5$ and/or $\rho$, ${\cal B}(a,\rho,N)$ increases rapidly and become prominent. The dependence of ${\cal B}(a,\rho,N)$ on $\rho$ is the reason that Eq. (\ref{eq:M}) allows $M_i^{\mathrm{(ii)}}$ to include the number of duplicated edges between any two nodes, meaning that Eq. (\ref{eq:B}) does not provide the quantitative aspect, but gives the qualitative prospect. Section \ref{subsec:clus-num} demonstrates good consistency between the analytical and numerical approaches.

\section{Numerical solutions}
\label{sec:num}
\subsection{Degree distribution}
\begin{figure}[ht]
	\includegraphics{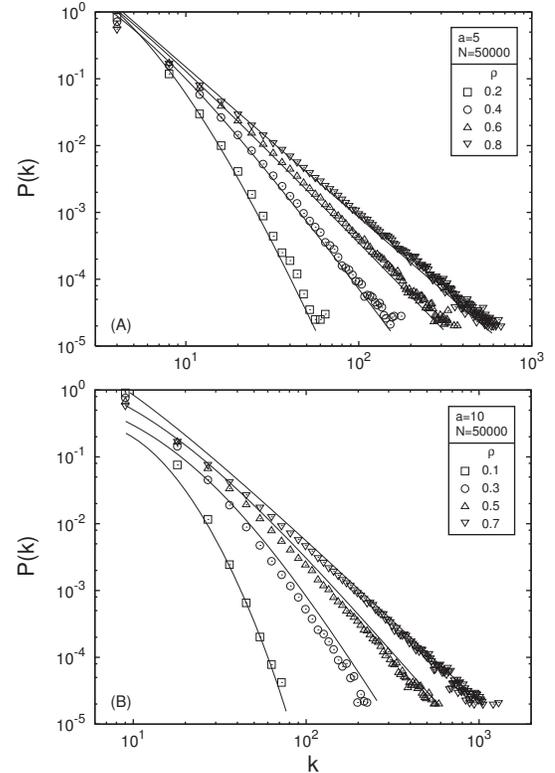}
	\caption{Degree distributions $P(k)$. Different symbols denote different numerical results and solid lines are depicted by using Eq. (\ref{eq:deg}) with $N=50 \ 000$. (A) $a=5$. (B) $a=10$.}
	\label{fig:deg}
\end{figure}

In order to confirm the analytical predictions, we performed numerical simulations of networks generated by using our model described in Secs. \ref{sec:model} and \ref{sec:ana}.  Figure \ref{fig:deg} (A) and \ref{fig:deg} (B) show degree distributions with different numerical conditions.  Solid lines come from Eq. (\ref{eq:deg}).  We show excellent agreement with the theoretical predictions.

\subsection{Clustering spectrum}
\label{subsec:clus-num}
\begin{figure}[ht]
\begin{center}
	\includegraphics{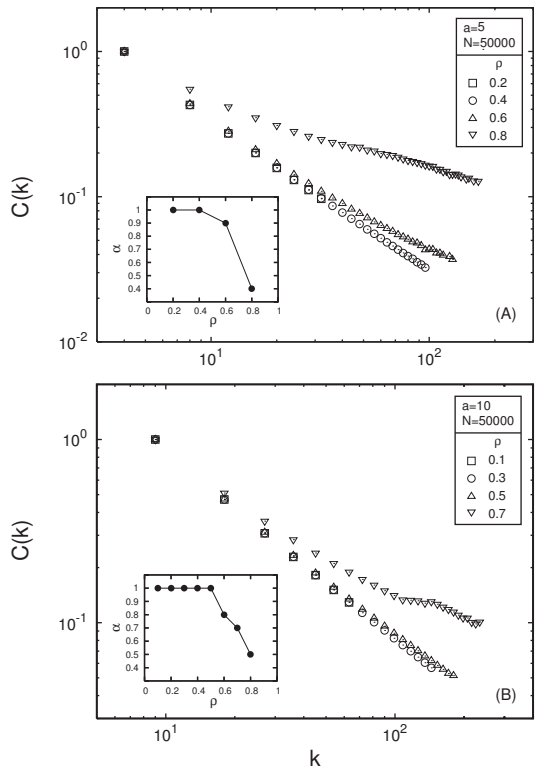}
\end{center}
	\caption{Clustering spectra $C(k)$. Different symbols denote different numerical conditions for $m$ with fixed $N=50 \ 000$. (A) $a=5$. (B) $a=10$. The insets show the relationship between $\rho$ and $\alpha$, where $\alpha$ is defined by the exponent from rational distribution $C(k)\sim k^{-\alpha}$}
	\label{fig:clus}
\end{figure}

Based on our model we obtain the degree-clustering coefficient correlations (clustering spectra) shown in Fig. \ref{fig:clus}.  For $\rho\leq 0.5$, the power-law (SF) regime is established as predicted in Eq. (\ref{eq:cluster}), indicating the hierarchical feature in generated complex networks.  For $\rho>0.5$, we obtain a gentle decay of $C(k)$ for larger $k$.  This decay can be explained by the tendency of ${\cal B}(a,\rho,N)$ as a function of $\rho$. With increasing $\rho$, ${\cal B}(a,\rho,N)$ increases because of more overlapping clique, leading to the transformation of the $C(k)$ tail from rapid to flat. The gentle decay of the tail corresponds to less chance of establishment of hierarchical structure with larger $\rho$.

\subsection{Average clustering coefficient}
\begin{figure}[ht]
\begin{center}
	\includegraphics{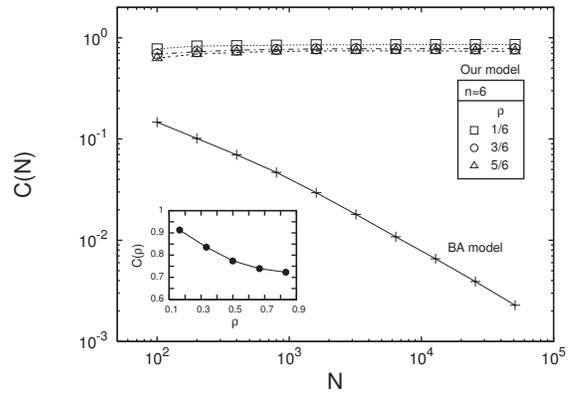}
\end{center}
	\caption{Comparison of average clustering coefficients $C(N)$ from two models.  $N$ varies from $100$ to $50 \ 000$ with fixed $a=6$. Inset: Dependence of $C$ on $\rho$ with $N=3 \ 000$ and $a=6$.}
	\label{fig:CN}
\end{figure}

In order to demonstrate that our model can construct a complex network with a high clustering coefficient with comparing the BA model, we numerically obtain average-clustering coefficients defined as $C(N)=(1/N)\sum_{i=1}^NC_i$.  Figure \ref{fig:CN}. shows both results from our model and the BA model and $C(N)$ from the BA model was predicted as  $C(N)\propto (\ln N)^2/N\simeq N^{-0.75}$ \cite{ref:CN,ref:rate}. In contrast, our model exhibits the independence of $C(N)$ on $N$ as well as higher $C(N)$ with different $\rho$. The feature has been reported to be found in the real-world networks \cite{ref:determin, ref:metabo-module} and is prominent property of hierarchical, small-world networks. The inset of Fig. \ref{fig:CN} shows the decay of $C$ as a function of $\rho$. As $\rho$ increases $C$ gently decreases, due to the increase of randomness in the network caused by more frequent overlapping.

\section{Conclusion}
\label{sec:conc}
We have proposed the growth network model with the merging clique mechanism.  The numerical simulations of the model have reproduced the statistical properties observed in the real-world network; power-law degree distributions with arbitrary exponent, power-law clustering spectrum, and average clustering coefficients are independent of network size.
 
In particular, we also have derived the analytical solution of the exponent following $\gamma=(\rho+1)/\rho$ by using a continuous approach via coarse-graining procedure.  The solution showed that the degree exponents are determined by the only ratio of the number of merging nodes to that of clique nodes and had excellent agreement with corresponding numerical simulations.

This relationship for $\gamma$ means that the degree exponent is controllable by tuning $\rho$ and implies that the real-world networks with decaying degree exponent tend to contain a large number of similar modules or communities with higher density as well as we may also be able to predict a degree exponent $\gamma$ when we can estimate the ratio.

In addition, our research expects that large-scale complex networks are consist of small scale classical networks, which have been called Erd\H{o}s-R\'enyi or regular graphs, suggesting that the classical graph theory is helpful for a making and analyzing the growing network model.  We hope that our model may become a bridging model between scale-free networks and classical networks.

Finally, because of successful reproduction of some remarkable characteristics that can be found in biological systems, our approach may become a useful tool to provide comprehensive aspects of and to disentangle evolutionary processes of self-organized biological networks and biocomplexity.

\appendix

\section{Degree correlation}
\label{sec:appendix}
To show proof of the assumption for the uncorrelated average nearest-neighbors degree in Sec. \ref{subsec:clus-ana}, we give analytical and numerical solutions for degree correlation of the model.

First, we introduce the analytical solutions. Degree correlation the represents the average degree of neighbors of node(s) with degree $k$ and is defined as
\begin{equation}
\bar{k}_{nn}(k)=\sum_{k'}k'P(k'|k),
\label{eq:0}
\end{equation}
where the conditional probability $P(k'|k)$ is the frequency that a node with degree $k$ connects to a node with degree $k'$. Using Kronecker's delta function, we redefined the degree correlation as
\begin{equation}
\bar{k}_{nn}(k)=\frac{\sum_{i=1}^{N}\langle k_{nn}\rangle_i\times \delta(k_i-k)}{\sum_{i=1}^N\delta(k_i-k)},
\label{eq:2.5}
\end{equation}
where $\langle k_{nn}\rangle_i$ denotes the average nearest-neighbor degree, written as
\begin{equation}
\langle k_{nn}\rangle_i=\frac{R_i}{k_i}.
\label{eq:2}
\end{equation}
In Eq. (\ref{eq:2}), $R_i(t)$ denotes the sum of the degree of neighbors of node $i$ and are represented as
\begin{equation}
R_i=\sum_{{h\in{\cal V}(i)}}k_h,
\label{eq:1}
\end{equation}
where ${\cal V}(i)$ correspond to the set of neighbors of node $i$. 

\begin{figure*}[t]
	\includegraphics{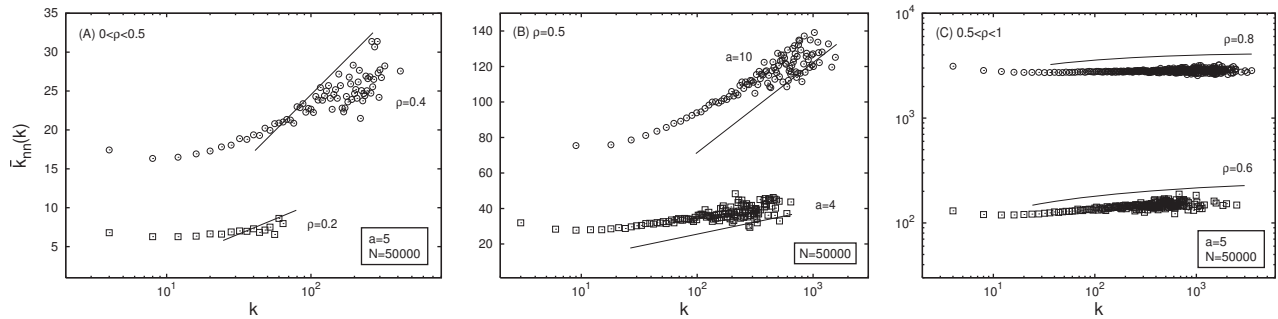}
	\caption{Degree correlations $\bar{k}_{nn}(k)$ with $N=50 \ 000$. (A) $0<\rho<0.5$. (B) $\rho=0.5$. (C) $0.5<\rho<1$. Symbols correspond to the numerical results and the solid lines are given by Eqs. (\ref{eq:7}), (\ref{eq:9}), and (\ref{eq:11}), respectively. Note that (A) and (B) are depicted by using single logarithmic plot, and (C) is a double logarithmic plot.}
	\label{fig:knn}
\end{figure*}

To get the analytical solution for the time evolution of $R_i(t)$ by means of the the rate equation approach \cite{ref:rate}, we clarify the two conditions that contribute to increase of rates in the equation.  The conditions are
\begin{enumerate}[i)]
\item Node $i$ is selected by the preferential attachment (PA) rule. 
\item Neighbor(s) of node $i$ are chosen by the PA rule. 
\end{enumerate}
Therefore summation of these contributions can be expressed as
\begin{eqnarray}
\frac{\diff R_i(t)}{\diff t}=&&m\Pi_i[(a-m)(a-1)+(m-1){\cal K}(t)] \nonumber \\
&&+\sum_{h\in{\cal V}(i)}m\Pi_h(a-1).
\label{eq:3}
\end{eqnarray}
The first and the second term are derive from condition i) and condition ii), respectively. In Eq. (\ref{eq:3}), the ${\cal K}(t)$ expectation values of the degree of a node with the PA rule, written as Eq. (\ref{eq:4}), are approximated as Eq. (\ref{eq:5}) for the solution
\begin{eqnarray}
{\cal K}(t)&=&\sum_{h=1}^{t}k_h\Pi_h\label{eq:4}\\
&\simeq&\frac{1}{a(a-1)t}\int_1^t\left[{a \choose 2}\left(\frac{t}{x}\right)^{\rho}\right]^2\diff x.
\label{eq:5}
\end{eqnarray}
The degree depends on time $t$ as indicated,
\begin{equation}
k_i(t)\simeq{a \choose 2}\left(\frac{t}{s}\right)^{\rho},
\label{eq:5.5}
\end{equation}
derived in Sec. \ref{subsec:deg-ana}. Then we get
\begin{equation} 
{\cal K}(t) \simeq
\left\{
\begin{array}{ll}
\frac{a(a-1)}{4(1-2\rho)}=\mathrm{const} & (0<\rho<0.5) \\
\frac{a(a-1)}{4}\ln t & (\rho=0.5) \\
\frac{a(a-1)}{4(2\rho-1)}t^{2\rho-1} & (0.5<\rho<1) \\
\end{array}\label{eq:K}
\right.
\end{equation}
which proves the behaviors the depend on $\rho$. Furthermore, the initial condition at time $s$ is the sum of degrees of the others of node $i$ in the clique and is given by
\begin{equation}
R_i(s)=(a-m-1)(a-1)+m{\cal K}(s).
\end{equation}

Then, we show the solutions as a function of $\rho$ as follows.

(i) $0<\rho<0.5$. In the range, inserting Eq. (\ref{eq:K}) into Eq. (\ref{eq:3}), the sum $R_i(t)$ is described as
\begin{equation}
R_i(t)\simeq\frac{mA_0}{a(a-1)}{a \choose 2}\left(\frac{t}{s}\right)^{\rho}\ln\frac{t}{s},
\label{eq:6}
\end{equation}
where $A_0=(a-m-1)(a-1)+m[a(a-1)/4(1-2\rho)]$. Substituting this into Eq. (\ref{eq:2}), the dominant behavior of the degree correlation is given by
\begin{equation}
\bar{k}_{nn}(k)\simeq\frac{A_0}{a-1}\ln\frac{2k}{a(a-1)},
\label{eq:7}
\end{equation}
and represents weak assortativity, defined as positive correlation between degree $k$ and the average nearest-neighbor degree of the node with degree $k$. Moreover, the tendency of the assortativity is determined by $a$ and/or $\rho$.

(ii) $\rho=0.5$. In this case, inserting Eq. (\ref{eq:K}) into Eq. (\ref{eq:3}), the dominant structure of the
sum $R_i(t)$ is
\begin{eqnarray}
R_i(t)\simeq&&\frac{m(m-1)}{8}{a \choose 2}\left(\frac{t}{s}\right)^{\rho}\ln\left(\frac{t}{s}\right)^{\rho}\ln t \nonumber \\
&&+m\frac{a(a-1)}{4}\left(\frac{t}{s}\right)^{\rho}\ln t.
\label{eq:8}
\end{eqnarray}
The same step can be applied as shown in case (i), the dominant factors for the correlation function become
\begin{equation}
\bar{k}_{nn}(k)\simeq\left[\frac{a(a-2)}{32}\ln\frac{2k}{a(a-1)}+\frac{a}{4}\right]\ln N_G,
\label{eq:9}
\end{equation}
where $N_G=N/(a-m)$ and also has weak assortativity as seen in case (i) when $a>2$. In the case of $a=2$, the model is the same as the BA model, then the degree correlation is described as $\bar{k}_{nn}(k)=(\ln N)/2$ in common with the reported results \cite{ref:knn,ref:rate}.

(iii) $0.5<\rho<1$. In this case, inserting Eq. (\ref{eq:K}) into Eq. (\ref{eq:3}), the sum $R_i(t)$ is
\begin{widetext}
\begin{equation}
R_i(t)\simeq\frac{m(m-1)}{4(2\rho-1)^2}{a \choose 2}\left[\left(\frac{t}{s}\right)^{\rho}t^{2\rho-1}-s^{\rho-1}t^{\rho}\right]+m\frac{a(a-1)}{4(2\rho-1)}s^{\rho-1}t^{\rho}.
\label{eq:10}
\end{equation}
To obtain the dominant factors of the correlation function we take the same procedure as described above,
\begin{equation}
\bar{k}_{nn}(k)\simeq\frac{m(m-1)}{4(2\rho-1)^2}\left[1-\frac{m+1-4\rho}{m-1}\left\{\frac{2k}{a(a-1)}\right\}^{1/\rho-2}\right]N_G^{2\rho-1}
\label{eq:11}
\end{equation}
indicates the uncorrelated feature for the large $k$.
\end{widetext}

Second, we give numerical solutions of the correlation of the model. For $\rho\leq 0.5$ [see Fig. \ref{fig:knn} (A) and \ref{fig:knn} (B)], average nearest-neighbor degrees of nodes with degree $k$ grow logarithmically with increasing $k$, and the uncorrelated features are shown for small $a$ and/or $\rho$. The solid lines correspond to the analytical results described above, and are fitted to the numerical ones for large $k$.  For $\rho>0.5$ [see Fig. \ref{fig:knn} (C)], the average degrees have no correlations with respect to degree $k$. As in the previous case, the analytical results are represented as the solid lines are fitted to numerical ones. On the basis of these analytical and numerical evidences, we assume the degree correlation is uncorrelated in Sec. \ref{subsec:clus-ana}.


\end{document}